# PERFORMANCE ANALYSIS OF DECISION DIRECTED MAXIMUM LIKELIHOOD MIMO CHANNEL TRACKING ALGORITHM

Ebrahim Karami

Centre for Wireless Communications (CWC), University of Oulu,

P.O. Box 4500, FIN-90014, Oulu, Finland

Abstract— In this paper, the performance of decision directed (DD) maximum likelihood (ML) channel tracking algorithm is analyzed. The ML channel tracking algorithm presents efficient performance especially in the decision directed mode of the operation. In this paper, after introducing the method for analysis of DD algorithms, the performance of ML Multiple-Input Multiple-Output (MIMO) channel tracking algorithm in the DD mode of operation is analyzed. In this method channel tracking error is evaluated for given decision error rate. Then, the decision error rate is approximated for given channel tracking error. By solving these two derived equations jointly, both the decision error rate and the channel tracking error are computed.

The presented analysis is compared with simulation results for different channel ranks, Doppler frequency shifts, and SNRs; and it is shown that the analysis is a good match for simulation results especially in high rank MIMO channels and high Doppler shifts.

## I. INTRODUCTION

In the recent years, the use of multiple antennas has been developed to increase spectral efficiency and diversity gain in wireless communications systems [1]. Use of Multiple-Input Multiple-Output (MIMO) channels, i.e. when multiple antennas are used in both receiver and transmitter, has much higher spectral efficiency versus Single-Input Single-Output (SISO), Single-Input Multiple-Output (SIMO), and Multiple-Input Single-Output (MISO) channels [2]. Moreover, the diversity gain of MIMO channels is nearly of second order when channel matrix has full rank which is much higher than when antenna array is used in one side of the communication's link. Therefore, by employing MIMO channels both cutoff and average capacities are improved and the mobility of wireless communications increased.

Use of channel estimation or equalization is vital in MIMO channels, whether channel is flat or frequency selective, unlike SISO channels where equalization is required only in frequency selective channels. MIMO detection algorithms can be divided into two main groups namely, direct or indirect. In direct detection algorithm, detection is done by a linear or nonlinear filter called an equalizer. But in indirect detector, first the channel state is estimated and then the data is detected using the estimated channel state. Therefore, when the indirect detection algorithm is used in MIMO channels, a joint detection unit is required, without which inter sub-stream interference occurs. Joint detection algorithms used in MIMO channels are developed based on Multi-User Detection (MUD) algorithms in CDMA systems. This is an outcome of the similarity between MIMO channels and CDMA systems formulation [3]. The only difference in MIMO channels is the spatial spreading instead of spreading using codes. Maximum Likelihood (ML) is the optimum joint detection algorithm [4]. The computational complexity of the optimum receiver is impracticable if the number of transmitting sub-streams is large [5]. On the other hand, with inaccurate channel information occurring when channel estimator tracking speed is insufficient for accurate tracking of the channel variations, the implementation of the optimum receiver is more complex. Therefore, the sub-optimum joint detection algorithm seems to be a more efficient solution. In this paper, the Minimum Mean Square Error (MMSE) detector which is the best linear joint detection algorithm due to its reasonable complexity and providing soft output, is considered as the joint detector [6], [7]. Of course, the discussion can be extended further to other joint detection algorithms.

Channel estimators may or may not use the training sequence. Although, the distribution of training symbols in a block of data affects the performance of systems [8], but due to simplicity, it is conventional to use the training symbols in the first part of each block. If the training sequence is not used, the estimator is called the blind channel estimator. A blind channel estimator uses information latent in statistical properties of the transmitting data [9]. The derivation of statistical properties of data can be done as direct or indirect. The scope of indirect blind methods are based on soft [10] or hard [11] decision directed algorithms using the previous estimation of the channel for detection of data and applying it for estimation of the channel in the last snapshot. Therefore, with decision directing, most of the non-blind algorithms can be implemented as blind. In MIMO channels, use of an initial training data is mandatory and without it channel estimator does not converge. Most of the channel estimation algorithms are designed for quasi static channels where the channel state can be considered unvarying over a block of data [11]. But as the channel state is changed over a block of data symbol by symbol, a tracking algorithm is required to update the estimated channel state. One of the most efficient MIMO channel tracking algorithms is the recursive least squares (RLS) based channel estimator. This algorithm presents efficient performance in the DD mode of operation [12-16]. When the exact channel variation model is known at the receiver, more efficient algorithms can be designed. One of the most well-known tracking algorithms is the Kalman filtering estimation proposed by Kominankis et. al. [17, 18]. In this paper, a Kalman filter is used as a MIMO channel tracker. The performance of this algorithm is relatively acceptable for Rice channels where a part of the channel, due to line-of-sight components, is known. But this algorithm has high complexity in the order of 5. In [19], the maximum likelihood estimator is proposed for tracking of MIMO channels. This algorithm extracts equations for maximum likelihood estimation of a time-invariant channel and extends it to a time-variant channel. As such, it does not have a desirable performance for time-varying channels. In [20], the maximum likelihood algorithm with an efficient tracking performance is derived for time-varying MIMO channels. The ML MIMO channel tracking algorithm presents the best tracking performance with low complexity which is only in the order of 2. In [21], the performance of ML algorithm is analyzed in the training mode where transmitted symbols are known at the receiver.

Channel estimation error affects the performance of wireless systems as a multiplicative noise. Therefore when an estimate of channel estimation error is available at transmitter or receiver sides, transceiver optimization techniques can be applied to enhance the overall performance of the system. For example when we have this extra knowledge at the transmitter, we can optimize signaling constellation to increase the throughput [22, 23]. In a DD based channel estimation algorithm, the MSE of channel tracking is varying during tracking process and therefore to optimize the overall throughput, we need to update the estimated channel tracking error.

In this paper, a technique for analysis of the DD based algorithms is proposed. This technique is applied for performance analysis of the DD based blind ML MIMO channel tracking algorithm but it is applicable for all decision directed based algorithms. Analysis of decision directed algorithms open a new scope for modification of this class of algorithms with partial compensation of decision error.

The rest of this paper is organized as follows. In Section II, the signal transmission model and the channel model are briefly introduced. The model used is the same as in [21]. In Section III which is the main body of the paper, first, the proposed technique for the analysis of DD based algorithms is introduced, then the MSE of tracking is derived for the known decision error rate, and finally the decision error rate is calculated versus the SNR considering MMSE detection algorithm. Simulation results of the proposed receiver are presented and compared with the Kalman filtering approach in Section IV and finally Section V concludes the paper.

## II. PROBLEM DEFINITION

### A. The Channel Model

Block diagram of the transmitter in a spatial multiplexed MIMO system with $M$ antennas, with the same configuration as in [21], is shown in Fig. 1. The main input block is de-multiplexed to $M$ sub-blocks. Then all $M$ sub-blocks are transmitted separately via transmitters. In the receiver, as shown in Fig. 2, linear combination of all transmitted sub-blocks are distorted by time-varying Rayleigh or Ricean fades, where the Inter Symbol Interference (ISI) is observed under additive white Gaussian noise. In this paper, without loss of generality, flat fading MIMO channel with Rayleigh distribution under first order Markov model variation is assumed. The observable signal $r_k^i$ from receiver i (with i =1, …, $N$) at discrete time index k is

$$r_k^i = \sum_{j=1}^{M} h_k^{i,j} s_k^j + w_k^i, \qquad (1)$$

where $s_k^j$ is the transmitted symbol in time index k, $w_k^i$ is the additive white Gaussian noise in the ith received element, and $h_k^{i,j}$ is the propagation attenuation between the jth input and the ith output of the MIMO channel that is a complex number with Rayleigh distributed envelope. Therefore, in each time instance, $MN$ channel parameters must be estimated which severely vary in the duration of data block transmission with the following autocorrelation [24]

$$E\left\{ h_k^{i,j} \left[ h_l^{i,j} \right]^* \right\} \cong J_0\left( 2\pi f_D^{i,j} T |k - l| \right), \qquad (2)$$

where, $J_0(.)$ is the zero-order Bessel function of the first kind, superscript * denotes the complex conjugate, $f_D^{i,j}$ is Doppler frequency shift for the mth path between the jth transmitter and the ith receiver, and T is the duration of each symbol. According to the Wide Sense Stationary Uncorrelated Scattering (WSSUS) model of Bello [24], all the channel taps are independent, namely all $h_k^{i,j}$s vary independently according to the autocorrelation model of (2). Channel variations can be modeled by the following first order autoregressive (AR) model

$$h_k^{i,j} = \alpha_{i,j} h_{k-1}^{i,j} + v_{i,j,k}, \qquad (3)$$

where $\alpha_{i,j}$ is the coefficient of AR model and $v_{i,j,k}$s are zero-mean i.i.d. complex Gaussian processes with variances given by

$$E\left(v_{i,j,k}\left[v_{i,j,k}\right]^{*}\right)=\sigma^2_{v_{i,j,k}}. \tag{4}$$

Optimum $\alpha_{i,j}$ is derived by solving Wiener equations considering correlation function (2), as follows,

$$\alpha_{i,j} = J_0\left(2\pi f_D^{i,j} T\right). \tag{5}$$

A reasonable assumption, conventional in the most scenarios, is the equal Doppler shifts for all channel matrix elements, i.e. $f_D^{i,j} = f_D$. This assumption does not make any changes in the derived algorithm. Equations (1) can be rewritten in a matrix form as

$$r_k = H_k s_k + w_k, \tag{6}$$

where $r_k$ is the received vector, $H_k$ is the channel matrix, and $s_k$ is the transmitted symbol, all in time index $k$, and $w_k$ is the vector with i.i.d. AWGN elements with variance $\sigma_w^2$; also (3) can be rewritten in matrix form as

$$H_k = \alpha\, H_{k-1} + V_k, \tag{7}$$

where $V_k$ is a matrix with i.i.d. Rayleigh elements with variance $\sigma_V^2$, and $\alpha$ is a constant parameter that can be calculated by solving Wiener equation as follows

$$\alpha = J_0\left(2\pi f_D T\right). \tag{8}$$

It is obvious that larger Doppler rates lead to smaller $\alpha$ s and, therefore, faster channel variations. Because of the orthogonality between the channel state and the additive random part in first order AR channel model, the power of time-varying part of each tap is

$$P_k = E\left|h_{m,k}^{i,j}\right|^2 = \frac{\sigma_V^2}{1-\alpha^2}. \tag{9}$$

**B. The ML MIMO Channel Tracking Algorithm**

When MIMO channel is estimated and tracked, the following steps are performed for each snapshot [21].

Step 1. Initialization: $\hat{\boldsymbol{H}}_k = \boldsymbol{0}_{N\times M}$ and $E_0 = 1$, where $\boldsymbol{0}_{N\times M}$ is an $N\times M$ zero matrix and $E_k$ is the variance of channel tracking error when the data is known (for example in training mode) in time index k.

Step 2. If the algorithm is in the DD mode the data vector is estimated using

$$\hat{s}_k = g\left[\left(\hat{\boldsymbol{H}}_{k-1}^H \hat{\boldsymbol{H}}_{k-1} + \sigma_w^2 I_M\right)^{-1} \hat{\boldsymbol{H}}_{k-1}^H r_k\right], \quad (10)$$

where $g(.)$ is a function modeling the decision device which is the signum function in the special case of BPSK signaling and superscript $H$ denotes the conjugate transpose.

Step 3. Calculation of $\beta_k$ using

$$\beta_k = \frac{\alpha^2 E_{k-1} + \sigma_V^2}{\sigma_w^2 + M\left(\alpha^2 E_{k-1} + \sigma_V^2\right)}. \quad (11)$$

Step 4. Updating the channel vector estimate using

$$\hat{\boldsymbol{H}}_k = \alpha \hat{\boldsymbol{H}}_{k-1} + \beta_k \left(r_k - \alpha \hat{\boldsymbol{H}}_{k-1} \hat{s}_k\right) \hat{s}_k^H, \quad (12)$$

where $\hat{s}_k$ is multiplexed between data vector and its estimated value in the training mode and the DD mode, respectively.

Step 5. Updating channel $E_k$, using

$$E_k = \left\{1 - \frac{\alpha^2 E_{k-1} + \sigma_V^2}{\sigma_w^2 + M\left(\alpha^2 E_{k-1} + \sigma_V^2\right)}\right\} \left(\alpha^2 E_{k-1} + \sigma_V^2\right). \quad (13)$$

Step 6. Return to Step 2 for the next snapshot.

C. The Outputs of the Analysis

The following items are estimated by the analysis presented in this paper.

1. Decision (Bit) Error Rate is defined as

$$P_{e_k} = P(\hat{s}_k \neq s_k). \quad (14)$$

2. MSE of tracking defined as

$$\sigma_{\mathbf{H}_k}^2 = \frac{1}{MN} trace\{E(\xi_k^H \xi_k)\}, \qquad (15)$$

where $\xi_k$ is the channel estimation error and defined as follows

$$\xi_k = \hat{\mathbf{H}}_k - \mathbf{H}_k. \qquad (16)$$

It must be noted that when the algorithm operates in the training based mode, $\sigma_{\mathbf{H}_k}^2$ is equal to $E_k$ as defined previously.

## III. THE ANALYSIS OF THE ALGORITHM

In this Section, the analysis of the DD ML MIMO channel tracking algorithm is presented. The Section has 4 parts. In the first part, the analysis technique is introduced. Then in the following parts of the analysis the derivation of the MSE of tracking and the decision error rate are presented, and finally the flowchart of the analysis is introduced.

### A. Analysis Technique

In this part, the structure of the analysis method is presented. In this method, first the channel tracking error (output SNR for equalizers) is calculated for a known decision error rate as follows

$$\sigma_{est}^2 = f(p_e, \sigma_w^2), \qquad (17)$$

where $\sigma_{est}^2$ is variance of channel tracking error, $\sigma_w^2$ is noise variance, $P_e$ is decision error rate, and $f(.,.)$ is a two-variable function.

Moreover, the decision error rate is calculated with $g(.,.)$ which is a two-variable function with channel tracking error and noise variance as its variables. Of course, this function completely depends on the detection algorithm,

$$P_e = g(\sigma_{est}^2, \sigma_w^2). \qquad (18)$$

Finally, by solving these two equations jointly, the decision error rate and the channel tracking error are computed in terms of noise variance given in the following equations

$$\sigma_{est}^2 = h_1(\sigma_w^2) \tag{19}$$

and

$$P_e = h_2(\sigma_w^2), \tag{20}$$

where $h_1(.)$ and $h_2(.)$ are non-linear functions of the noise variance.

**B. Computation of the Channel Tracking Error**

For calculation of the channel tracking error (15) and (16) are combined as follows,

$$\sigma_{\tilde{H}_k}^2 = \frac{1}{MN} trace\{E(\xi_k^H \xi_k)\} = \frac{1}{MN} trace\{E(\hat{H}_k - H_k)^H (\hat{H}_k - H_k)\}. \tag{21}$$

By expanding (21), we have,

$$\sigma_{\tilde{H}_k}^2 = \frac{1}{MN} trace\{E(\hat{H}_k^H \hat{H}_k - H_k^H \hat{H}_k - \hat{H}_k^H H_k + H_k^H H_k)\} = q_k - p_k - p_k^* + 1, \tag{22}$$

where $p_k$ and $q_k$ are two scalar parameters defined as follows,

$$p_k = \frac{1}{MN} trace\{E(H_k^H \hat{H}_k)\} \tag{23}$$

$$q_k = \frac{1}{MN} trace\{E(\hat{H}_k^H \hat{H}_k)\}. \tag{24}$$

$q_k$ is the average power of the channel matrix elements and, consequently, is real; and $p_k$ is the bias factor of estimation as proved in appendix A. (23) and (24) are calculated as follows (see appendices B and C)

$$p_k = \alpha^2(1-\beta_k)p_{k-1} + \beta_k(1-2P_{e_k}), \tag{25}$$

$$q_k = \alpha^2(1-2\beta_k + M\beta_k^2)q_{k-1} + 2\alpha^2 \beta_k p_{k-1}(1-M\beta_k)(1-2P_{e_k}) + \beta_k^2(M+\sigma_w^2). \tag{26}$$

Consequently, with the known decision error rate, the channel tracking error is calculated recursively after recursive calculation of $p_k$ and $q_k$ from (25) and (26) and their substitution in (22).

## C. Calculation of the Decision Error Rate

In this Section, an approximate equation for the average decision error rate and the channel tracking error is presented. Although there is no closed form expression for the BER of the MIMO joint detection algorithm but the proposed technique provides relatively precise approximation for the BER whose validation is justified through various simulations. It is obvious that this equation completely depends on the detection algorithm. In this paper, results are presented for the MMSE detector which is the best linear algorithm with relatively low complexity and soft outputs applied directly to a soft decoder when coded signals are transmitted. This procedure can be extended for other detectors. First the channel tracking error is added to noise as follows,

$$r_k = H_k s_k + w_k = \hat{H}_k s_k + w^c_k, \tag{27}$$

where $w^c_k$ is the effective noise vector in time index $k$ which is the sum of noise and channel tracking error where the variance of its elements can be calculated through the following equation,

$$\sigma_{w^c_k}^2 = \sigma_w^2 + M \sigma_{H_k}^2. \tag{28}$$

The distribution of the elements of the effective noise vector, $w^c_k$, is assumed to be Gaussian. Of course, as various simulations show, although this assumption is weak in the beginning of the operation in the DD mode but its approximation becomes more and more precise after a few snapshot operations in the DD mode.

In [25], the spectral efficiency of a CDMA system with a few of joint detection algorithms such as the MMSE is presented. The structure of a CDMA system is similar to a MIMO channel where only the spreading sequences in CDMA systems are substituted with spatial signature waveforms i.e. columns of the channel matrix. Using the spectral efficiency equation presented in [25], spectral efficiency of a MIMO channel with MMSE detector can be asymptotically approximated with the following equation

$$C = \frac{\kappa}{2} \log\left(1 + \gamma_k - \frac{1}{4}\left[\sqrt{\gamma_k\left(1+\sqrt{\kappa}\right)^2 + 1} - \sqrt{\gamma_k\left(1-\sqrt{\kappa}\right)^2 + 1}\right]^2\right), \tag{29}$$

where $\kappa$ and $\gamma_k$ are respectively load factors and SNR defined as follows,

$$\kappa = \frac{M}{N} \tag{30}$$

and

$$\gamma_k = \frac{1}{\sigma_{w^c_k}^2}. \tag{31}$$

We approximate $M \times N$ MIMO channel with $M$ independen SISO channel with the same spectral efficiency and decision error rate. Then the old equivalent SNR is mapped to new one to have the same spectral efficiency for both $M \times N$ MIMO and SISO systems. But the capacity of the equivalent SISO channel is as follows,

$$C = \frac{\kappa}{2} \log(1 + \gamma_{eq,k}), \tag{32}$$

where $\gamma_{eq,k}$ is the equivalent SNR. Comparing (C.8) and (C.11), the new equivalent SNR is calculated as follows,

$$\gamma_{eq} = \gamma_k - \frac{1}{4}\left[\sqrt{\gamma_k\left(1+\sqrt{\kappa}\right)^2 + 1} - \sqrt{\gamma_k\left(1-\sqrt{\kappa}\right)^2 + 1}\right]^2. \tag{33}$$

Finally, assuming for instance the BPSK signaling, the decision error rate is calculated as bellow,

$$P_{e_k} = Q\left(\sqrt{\gamma_{eq,k}}\right). \tag{34}$$

The validity of the proposed BER estimation technique can be observed in Fig. 3. It can be seen that when the channel has full rank the estimated BER is very close to simulation result in a wide range of $E_b/N_0$ but when channel is in half rank the proposed technique has high precision only in low values of $E_b/N_0$.

## D. The Flowchart of the Analysis

The flowchart for analysis of the ML MIMO channel tracking algorithm is shown in Fig. 4. As it can be seen from this Fig., after initialization of the algorithm, if the algorithm is in the training mode, the decision error rate is set to be zero, but if the algorithm is in the DD mode, at first the decision error rate is calculated by equation (28)-(34), then the calculated decision error rate is recursively applied to (25) and (26) for calculation of $p_k$ and $q_k$ and finally they are substituted in (22) for calculating the MSE of tracking.

## IV. SIMULATION RESULTS

In this Section, the analytical results are compared to the corresponding simulation results and the analytical results from [21] which ignores the effect of decision errors. In all cases 4 receiving antennas, 2 and 4 transmitting antennas, $f_D T$ equals to 0.004 and 0.01, and $E_b/N_0$ equal to 5dB are assumed and the decision error rate and the variance of the channel tracking error are considered as two comparison criteria. In all cases, the first 20 symbols of each block is considered as training and the other 180 symbols are considered as data symbols where the ML algorithm works blind over 180 symbols. The simulation results are averaged over 10000 independent simulations. In the rest of this Section the comparison for the BER and the channel tracking error is done in two separate parts.

## A. Comparison of the BER for Simulation and Analytical Results

The BER curves are plotted in Figs. 5 and 6 corresponding to $f_D T$ equal to 0.004 and 0.01, respectively. As shown in Fig. 5, when channel is in full rank i.e. 4×4 channel, the decision error rate derived from the analysis, completely matches with the simulation results. But in a half-rank channel, i.e. 2×4 channel, a noticeable difference up to 45 percent is observed between the simulation and the analytical results. This gap is result of inefficiency of the approximations applied in Section III.C for low decision error rates near error floor. MIMO channels always suffer from the error floor due to residual spatial multiplexing interference but the presented analysis in the Section III.C does not consider the exact distribution of this interference and therefore it presents better approximation for decision error rates above this floor.

When $f_DT = 0.01$, as shown in Fig. 6, in both half-rank and full rank channels, the difference observed between the simulation and the analytical results is negligible and the precision of analysis is relatively good. In this case since channel tracking error has main rule in the observed error floor, the effect of residual spatial multiplexing interference which affects the precision of the analysis is less than Fig. 5 and therefore we achieve the better precision for the presented analysis.

**B. Comparison of the MSE of Tracking for Simulation and Analytical Results**

The curves for MSE of tracking are shown in Figs. 7-10 where the first two Figs. are for $f_DT = 0.004$ and the last two Figs. are for $f_DT = 0.01$.

When the channel is in full rank and for $f_DT = 0.004$ as shown is Fig. 7, the presented analysis is very close to the simulation results. Although at the beginning of the blind mode, the analytical curve has a peak, but after a few symbols both curves are quite matched.

But for a half rank channel where $f_DT = 0.004$ as shown in Fig. 8, due to the small value of the decision error rate, 3 curves are very close together. In this case, simulation results are more matched with the analytical results than the analysis presented in [21]. In this case, the peak observed in the analytical results of 4×4 channel is not observed, but on the other hand the difference between simulation and analysis is increased in the blind mode.

For 4×4 channel where $f_DT = 0.01$ as shown in Fig. 9, the difference between simulation and analytical curves is much greater. The same as the $f_DT = 0.004$ case, a peak is observed in the initial part of the blind mode operation. And finally, when the channel is in half rank where $f_DT = 0.01$ as shown in Fig. 10, the result obtained is similar to $f_DT = 0.004$ where all three curves are very close.

## V. CONCLUSION

In this paper, the performance of the decision-directed maximum likelihood channel tracking algorithm is analyzed. After introducing a new analytical method for decision directed algorithms, the performance of the ML MIMO channel tracking algorithm in the decision directed mode of operation is analyzed. In this method, the channel tracking error is evaluated for a known decision error rate. Then, the decision error rate is computed for a known channel tracking error. Later, by solving these two derived equations jointly, both the decision error rate and the channel tracking error are computed. Of course, deriving the decision error rate for special multiplexed MIMO systems in closed form is impractical. Therefore, in this paper by a reasonable assumption the decision error rate is calculated via the spectral efficiency formula presented by S. Verdu et. al. for CDMA systems [25]. This technique provides a relatively good approximation for the decision error rate especially for the full rank MIMO channel whose validity is confirmed through various simulations.

The analytical results are compared to the corresponding simulation results and the efficiency of the presented analysis is confirm especially when the decision error rate is higher than the error floor introduced by residual spatial multiplexing interference in the MIMO channels.

## APPENDIX A. PROOF OF $p_k$ AS THE BIAS FACTOR OF ESTIMATION

The relationship between channel matrix and its estimate can be written as follows,

$$\hat{\boldsymbol{H}}_k = z_k \boldsymbol{H}_k + \boldsymbol{T}_k, \tag{A.1}$$

where $z_k$ is the bias factor of estimation and by solving Wiener equations, it can be calculated from the following equation,

$$z_k = \frac{E\{trace(\boldsymbol{H}_k^H \hat{\boldsymbol{H}}_k)\}}{E\{trace(\boldsymbol{H}_k^H \boldsymbol{H}_k)\}}, \tag{A.2}$$

where $\boldsymbol{T}_k$ is a random matrix with the following covariance matrix,

$$E\langle T_k^H T_k \rangle = E\langle \hat{H}_k^H (\hat{H}_k - p_k H_k) \rangle = E\langle \hat{H}_k^H \hat{H}_k \rangle - p_k E\langle \hat{H}_k^H H_k \rangle. \tag{A.3}$$

Considering channel matrix elements as i.i.d. unit variance elements (A.3) can be rewritten as follows,

$$z_k = \frac{1}{MN} trace\{E(H_k^H \hat{H}_k)\}. \tag{A.4}$$

By comparison of (A.4) and (7) it is proved that the $p_k$ parameter is the bias factor of the estimation.

APPENDIX B. RECURSIVE EQUATION FOR THE BIAS FACTOR $p_k$

In order to calculate the bias factor, both sides of (12) is multiplied by $H_k^H$ and then the expectation operator is applied as follows,

$$E(H_k^H \hat{H}_k) = \alpha E(H_k^H \hat{H}_{k-1}) + \beta_k E(H_k^H r_k \hat{s}_k^H) - \alpha \beta_k E(H_k^H \hat{H}_{k-1} \hat{s}_k \hat{s}_k^H). \tag{B.1}$$

Due to the independency of data with its estimation and noise, we have,

$$E(H_k^H \hat{H}_k) = \alpha E(H_k^H \hat{H}_{k-1})(1 - \beta_k E(\hat{s}_k \hat{s}_k^H)) + \beta_k E(H_k^H r_k \hat{s}_k^H) \tag{B.2}$$

and

$$p_k = \frac{\alpha}{MN} trace\{E(H_k^H \hat{H}_{k-1})(1 - \beta_k E(\hat{s}_k \hat{s}_k^H))\} + \frac{\beta_k}{MN} trace\{E(H_k^H r_k \hat{s}_k^H)\}. \tag{B.3}$$

The second expectation on the right side of (B.3) is the autocorrelation of the estimated data vector which is unity mainly due to the orthonormality of its elements. Therefore, (B.3) can be rewritten as follows,

$$p_k = \frac{\alpha(1 - \beta_k)}{MN} trace\{E(H_k^H \hat{H}_{k-1})\} + \frac{\beta_k}{MN} trace\{E(H_k^H r_k \hat{s}_k^H)\}. \tag{B.4}$$

The first term on the right side of (B.4) can be computed by applying the channel model (7) as follows,

$$trace\{E(H_k^H \hat{H}_{k-1})\} = trace\{E([\alpha \ H_{k-1} + V_k]^H \hat{H}_{k-1})\}. \tag{B.5}$$

And then

$$\text{trace}\{E(\boldsymbol{H}_k^H \hat{\boldsymbol{H}}_{k-1})\} = \alpha\, \text{trace}\{E(\boldsymbol{H}_{k-1}^H \hat{\boldsymbol{H}}_{k-1})\} + \text{trace}\{E(\boldsymbol{V}_k^H \hat{\boldsymbol{H}}_{k-1})\}. \tag{B.6}$$

The first term on the right side of (B.6) is proportional to the bias factor, and the second is the zero matrix due to the independency of $\boldsymbol{V}_k$ with $\hat{\boldsymbol{H}}_{k-1}$. Therefore, (B.6) can be simplified as follows,

$$\text{trace}\{E(\boldsymbol{H}_k^H \hat{\boldsymbol{H}}_{k-1})\} = \alpha M N p_{k-1}. \tag{B.7}$$

Then, the second term on the right hand side of (B.4) must be calculated. Therefore, $r_k$ is replaced with its equivalent using (6) as follows,

$$\text{trace}\{E(\boldsymbol{H}_k^H \boldsymbol{r}_k \hat{\boldsymbol{s}}_k^H)\} = \text{trace}\{E(\boldsymbol{H}_k^H [\boldsymbol{H}_k \boldsymbol{s}_k + \boldsymbol{w}_k]\hat{\boldsymbol{s}}_k^H)\}. \tag{B.8}$$

Then

$$\text{trace}\{E(\boldsymbol{H}_k^H \boldsymbol{r}_k \hat{\boldsymbol{s}}_k^H)\} = \text{trace}\{E(\boldsymbol{H}_k^H \boldsymbol{H}_k) E(\boldsymbol{s}_k \hat{\boldsymbol{s}}_k^H)\} + \text{trace}\{E(\boldsymbol{w}_k \hat{\boldsymbol{s}}_k^H)\}. \tag{B.9}$$

The last term on the right hand side of (B.9) is zero and due to the i.i.d. assumption for the elements of channel matrix, we have

$$E(\boldsymbol{H}_k^H \boldsymbol{H}_k) = N I_M. \tag{B.10}$$

Consequently, (B.8) is simplified as follows

$$\text{trace}\{E(\boldsymbol{H}_k^H \boldsymbol{r}_k \hat{\boldsymbol{s}}_k^H)\} = N\, \text{trace}\{E(\boldsymbol{s}_k \hat{\boldsymbol{s}}_k^H)\}. \tag{B.11}$$

Therefore,

$$\text{trace}\{E(\boldsymbol{H}_k^H \boldsymbol{r}_k \hat{\boldsymbol{s}}_k^H)\} = MN\, E(s_{k,i} \hat{s}_{k,i}^*), \tag{B.12}$$

where $s_{k,i}$ is an arbitrary element of the data vector and $\hat{s}_{k,i}$ is its estimation. Computation of $E(s_{k,i}\hat{s}_{k,i}^*)$ by assuming the known decision error rate and the type of modulation is very simple and straightforward and for BPSK modulated signal as follows,

$$\begin{aligned} E(s_{k,i}\hat{s}_{k,i}^*) = &\, P(s_{k,i} = +1, \hat{s}_{k,i} = +1) - P(s_{k,i} = +1, \hat{s}_{k,i} = -1) \\ &- P(s_{k,i} = -1, \hat{s}_{k,i} = +1) + P(s_{k,i} = -1, \hat{s}_{k,i} = -1). \end{aligned} \tag{B.13}$$

Using Bays' theorem, we have

$$E(s_{k,i}\hat{s}_{k,i}{}^*) = P(\hat{s}_{k,i}=+1|s_{k,i}=+1)P(s_{k,i}=+1) - P(\hat{s}_{k,i}=-1|s_{k,i}=+1)P(s_{k,i}=+1)$$
$$- P(\hat{s}_{k,i}=+1|s_{k,i}=-1)P(s_{k,i}=-1) + P(\hat{s}_{k,i}=-1|s_{k,i}=-1)P(s_{k,i}=-1) \quad \text{(B.14)}$$

And considering the following equations,

$$P(\hat{s}_{k,i}=+1|s_{k,i}=+1) = P(\hat{s}_{k,i}=-1|s_{k,i}=-1) = 1 - P_{e_k} \quad \text{(B.15)}$$

and

$$P(\hat{s}_{k,i}=-1|s_{k,i}=+1) = P(\hat{s}_{k,i}=+1|s_{k,i}=-1) = P_{e_k}. \quad \text{(B.16)}$$

Finally (B.13) is computed as follows,

$$E(s_{k,i}\hat{s}_{k,i}{}^*) = 1 - 2P_{e_k}. \quad \text{(B.17)}$$

Just as an instance and for general $2^m$ PSK modulation (B.17) can be easily extended to

$$E(s_{k,i}\hat{s}_{k,i}{}^*) = \prod_{t=0}^{m}\left(1 - 2P_{e_k}^t \sin(2^{t-1}\pi)^2\right). \quad \text{(B.18)}$$

where $P_{e_k}^t$ is probability of decision error for $t$th bit of PSK symbol $s_{k,i}$ and it is easy to see that (B.17) is an especial case of (B.18) when $m=1$.

Eventually by applying (B.3), (B.7), (B.12) and (B.17) in (B.4), the following recursive equation is derived for the bias factor,

$$p_k = \alpha^2(1-\beta_k)p_{k-1} + \beta_k(1-2P_{e_k}). \quad \text{(B.19)}$$

## APPENDIX C. RECURSIVE EQUATION FOR $q_k$

For calculation of $q_k$, first channel tracking algorithm (12) is rewritten as follows,

$$\hat{\boldsymbol{H}}_k = \alpha\hat{\boldsymbol{H}}_{k-1}(1-\beta_k \boldsymbol{s}_k \boldsymbol{s}_k^H) + \beta_k \boldsymbol{r}_k \boldsymbol{s}_k^H. \quad \text{(C.1)}$$

Then by applying (B.19) in (24) we have

$$q_k = \frac{1}{MN}\text{trace}\left\{E\left((\alpha\hat{\boldsymbol{H}}_{k-1}(1-\beta_k \hat{\boldsymbol{s}}_k \hat{\boldsymbol{s}}_k^H) + \beta_k \boldsymbol{r}_k \hat{\boldsymbol{s}}_k^H)^H (\alpha\hat{\boldsymbol{H}}_{k-1}(1-\beta_k \hat{\boldsymbol{s}}_k \hat{\boldsymbol{s}}_k^H) + \beta_k \boldsymbol{r}_k \hat{\boldsymbol{s}}_k^H)\right)\right\}. \quad \text{(C.2)}$$

Consequently, $q_k$ will be the summation of four components as follows

$$q_k = q_{k,1} + q_{k,2} + q_{k,3} + q_{k,4}, \tag{C.3}$$

where

$$q_{k,1} = \frac{\alpha^2}{MN} trace\left\{ E\left\langle \left(1 - \beta_k \hat{s}_k \hat{s}_k^H\right)^H \hat{H}_{k-1}^H \hat{H}_{k-1} \left(1 - \beta_k \hat{s}_k \hat{s}_k^H\right)\right\rangle \right\}, \tag{C.4}$$

$$q_{k,2} = \frac{\alpha \beta_k}{MN} trace\left\{ E\left\langle \left(1 - \beta_k \hat{s}_k \hat{s}_k^H\right)^H \hat{H}_{k-1}^H r_k \hat{s}_k^H \right\rangle \right\}, \tag{C.5}$$

$$q_{k,3} = \frac{\alpha \beta_k}{MN} trace\left\{ E\left\langle \hat{s}_k r_k^H \hat{H}_{k-1} \left(1 - \beta_k \hat{s}_k \hat{s}_k^H\right)\right\rangle \right\}, \tag{C.6}$$

$$q_{k,4} = \frac{\beta_k^2}{MN} trace\left\{ E\left\langle \hat{s}_k r_k^H r_k \hat{s}_k^H \right\rangle \right\}. \tag{C.7}$$

By applying the expectation operator on independent variables, (C.4) can be rewritten as follows,

$$q_{k,1} = \frac{\alpha^2}{MN} trace\left\{ E\left\langle \hat{H}_{k-1}^H \hat{H}_{k-1}\right\rangle E\left\langle \left(1 - \beta_k \hat{s}_k \hat{s}_k^H\right)\left(1 - \beta_k \hat{s}_k \hat{s}_k^H\right)\right\rangle\right\}. \tag{C.8}$$

Then,

$$q_{k,1} = \frac{\alpha^2}{MN}\left(1 - 2\beta_k + M\beta_k^2\right) trace\left\{ E\left\langle \hat{H}_{k-1}^H \hat{H}_{k-1}\right\rangle\right\} = \alpha^2\left(1 - 2\beta_k + M\beta_k^2\right) q_{k-1}. \tag{C.9}$$

Right hand side of (C.5) can be calculated as follows,

$$q_{k,2} = \frac{\alpha \beta_k}{MN} trace\left\{ E\left\langle \left(1 - \beta_k \hat{s}_k \hat{s}_k^H\right)^H \hat{H}_{k-1}^H \left(H_k s_k + w_k\right) \hat{s}_k^H \right\rangle \right\}. \tag{C.10}$$

Due to the orthogonality between noise and other variables, we have,

$$q_{k,2} = \frac{\alpha \beta_k}{MN} trace\left\{ E\left\langle s_k \hat{s}_k^H \left(1 - \beta_k \hat{s}_k \hat{s}_k^H\right)^H \right\rangle E\left\langle \hat{H}_{k-1}^H H_k \right\rangle \right\}. \tag{C.11}$$

By applying (B.7) and (B.17) in (C.11) we have,

$$q_{k,2} = \alpha^2 \beta_k p_{k-1}\left(1 - M\beta_k\right)\left(1 - 2P_{e_k}\right). \tag{C.12}$$

The right hand side of (C.6) is the conjugate of the right hand side of (C.5), therefore (C.6) is equivalent to (C.12) as follows,

$$q_{k,3} = q_{k,2}^* = \alpha^2 \beta_k p_{k-1}(1 - M\beta_k)(1 - 2P_{e_k}). \tag{C.13}$$

For computation of $q_{k,4}$ in (C.7), the received vector is replaced with its equivalent, from channel model equation (6), as follows,

$$q_{k,4} = \frac{\beta_k^2}{MN} trace\left\{E\left\langle \hat{s}_k (H_k s_k + w_k)^H (H_k s_k + w_k) \hat{s}_k^H \right\rangle\right\}. \tag{C.14}$$

Which after some manipulations, is computed as follows,

$$q_{k,4} = \beta_k^2 (M + \sigma_w^2) \tag{C.15}$$

And finally, after applying (C.9), (C.12), (C.13), and (C.15) in (C.3) we have

$$q_k = \alpha^2 (1 - 2\beta_k + M\beta_k^2) q_{k-1} + 2\alpha^2 \beta_k p_{k-1}(1 - M\beta_k)(1 - 2P_{e_k}) + \beta_k^2 (M + \sigma_w^2). \tag{C.16}$$

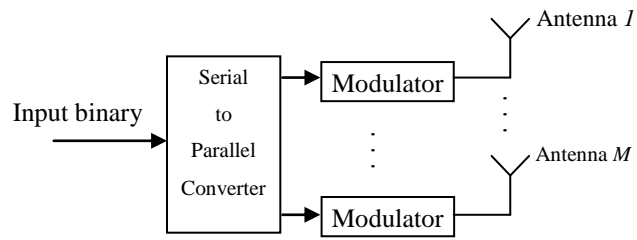

Fig. 1. Block diagram of a simple spatial multiplexed MIMO transmitter.

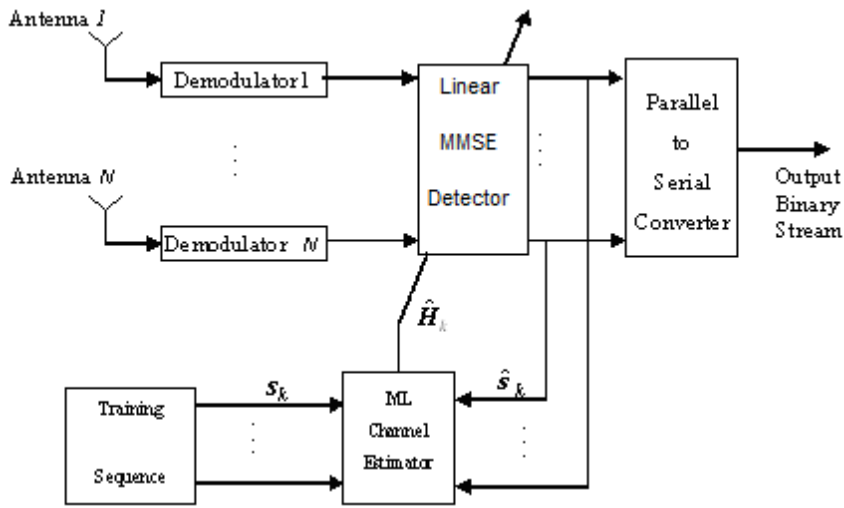

Fig. 2. Block diagram of the receiver.

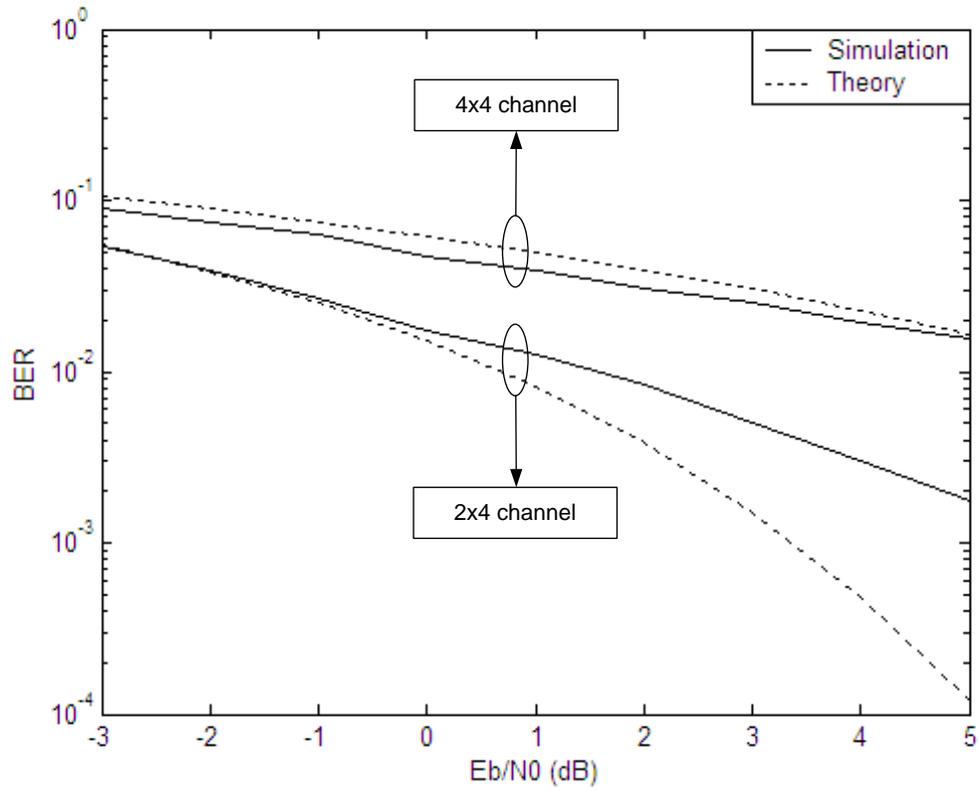

Fig. 3. Comparison of the BER of the MIMO channel derived by simulation and the proposed theory.

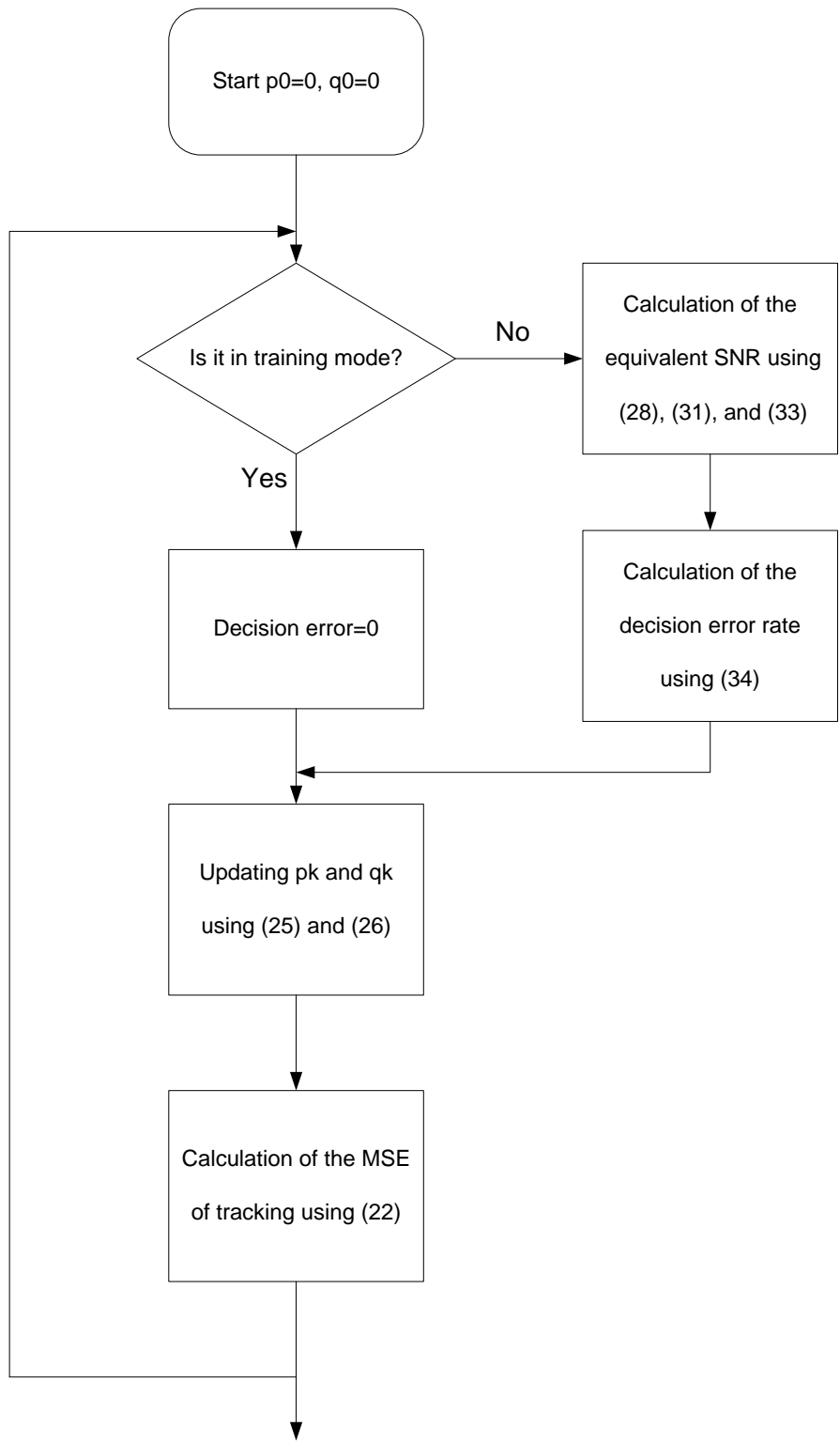

Fig. 4. Flowchart for analysis of the ML MIMO channel tracking algorithm in both training based and blind modes.

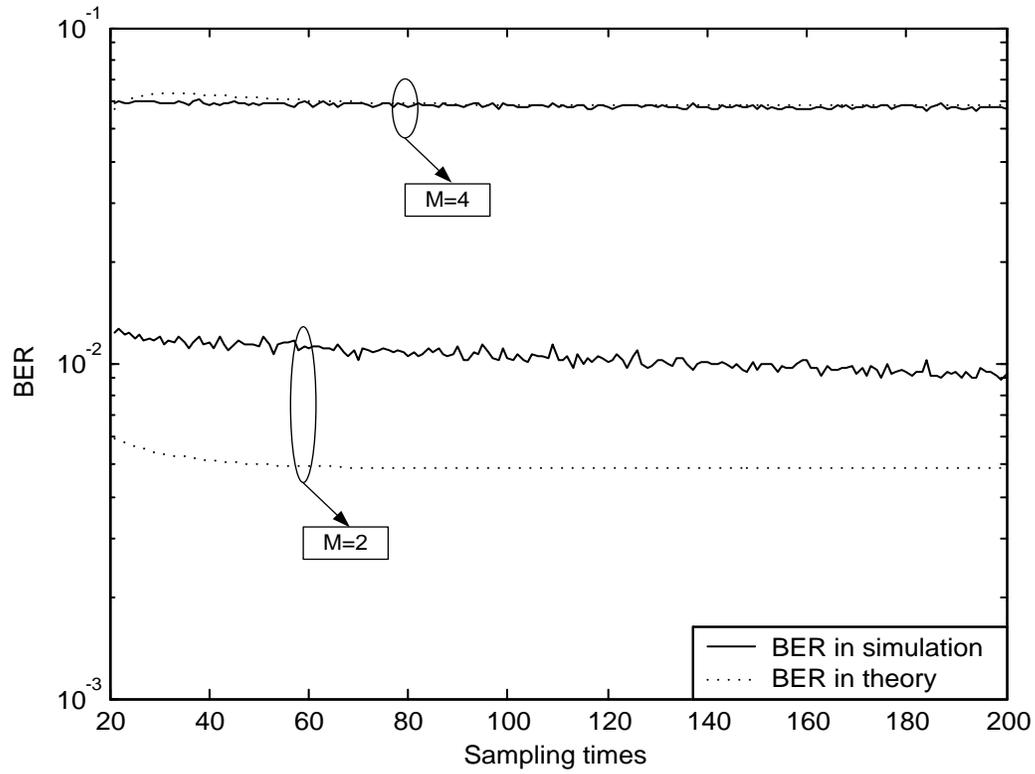

Fig. 5. Comparison of the BER in simulation and analytical results where $f_D T = 0.004$ and $E_b/N_0 = 5$dB.

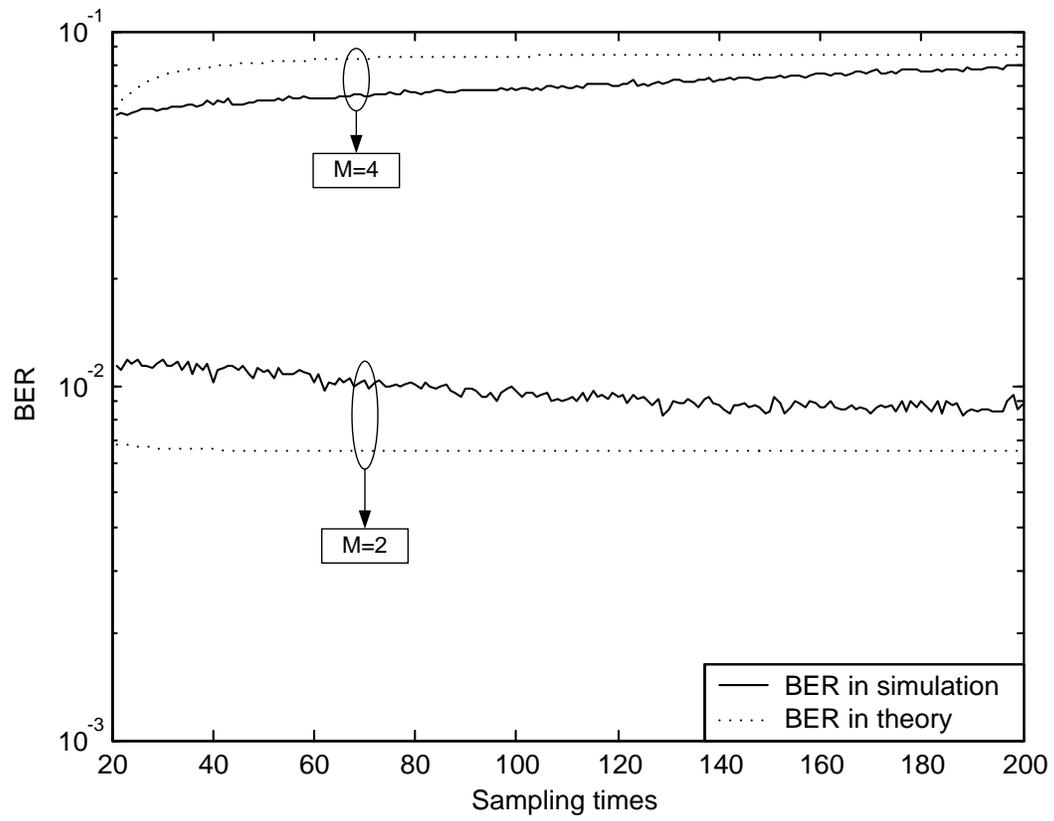

Fig. 6. Comparison of the BER in simulation and analytical results where $f_D T = 0.01$ and $E_b / N_0 = 5$dB.

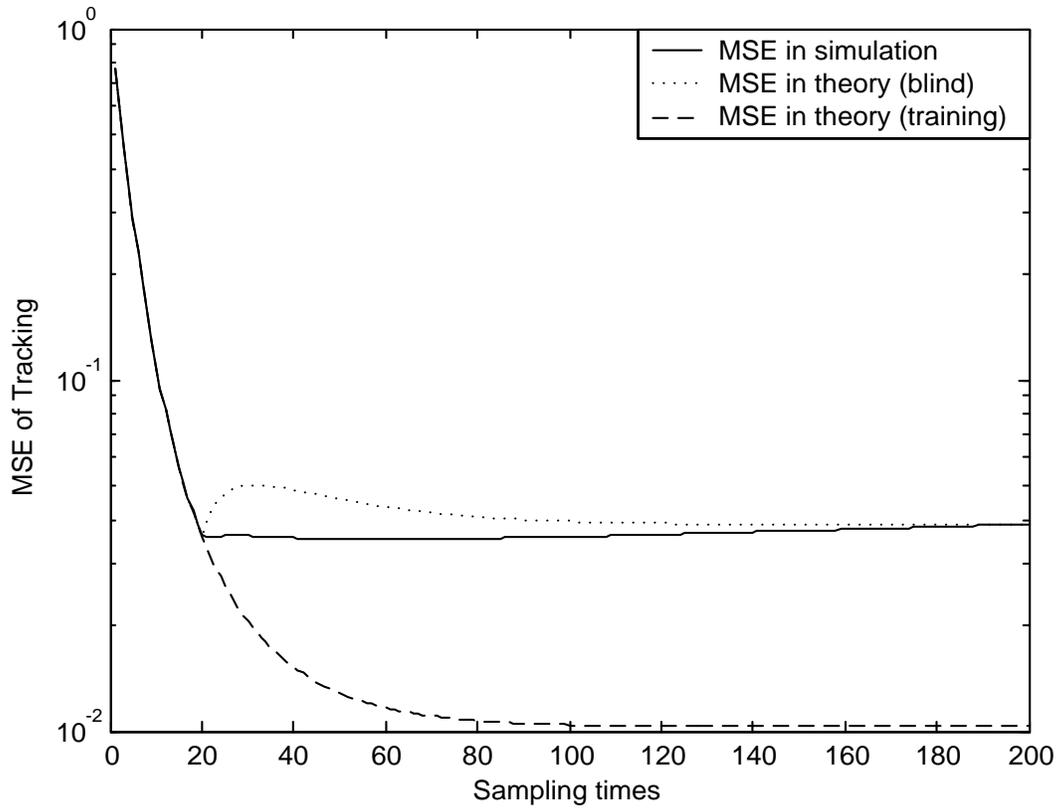

Fig. 7. Comparison of the MSE of Tracking in simulation and analysis where $f_D T$ =0.004 and 4×4 channel.

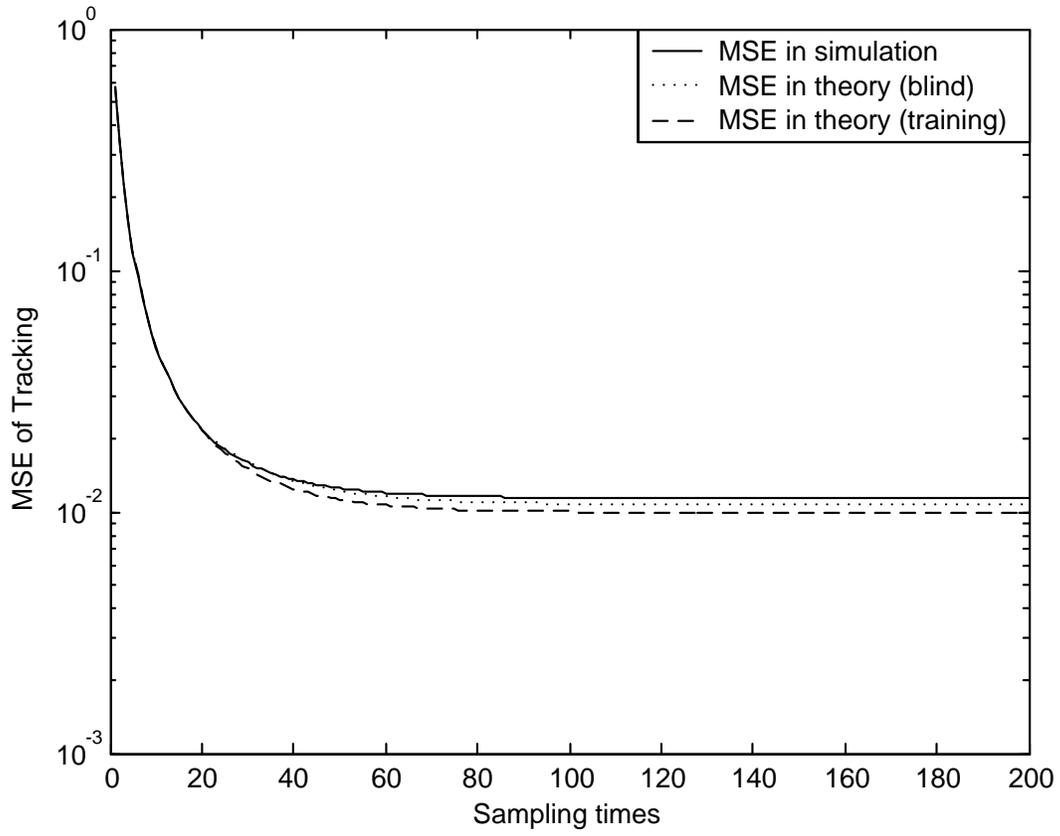

Fig. 8. Comparison of the MSE of Tracking in simulation and analysis where $f_D T$ =0.004 and 2×4 channel.

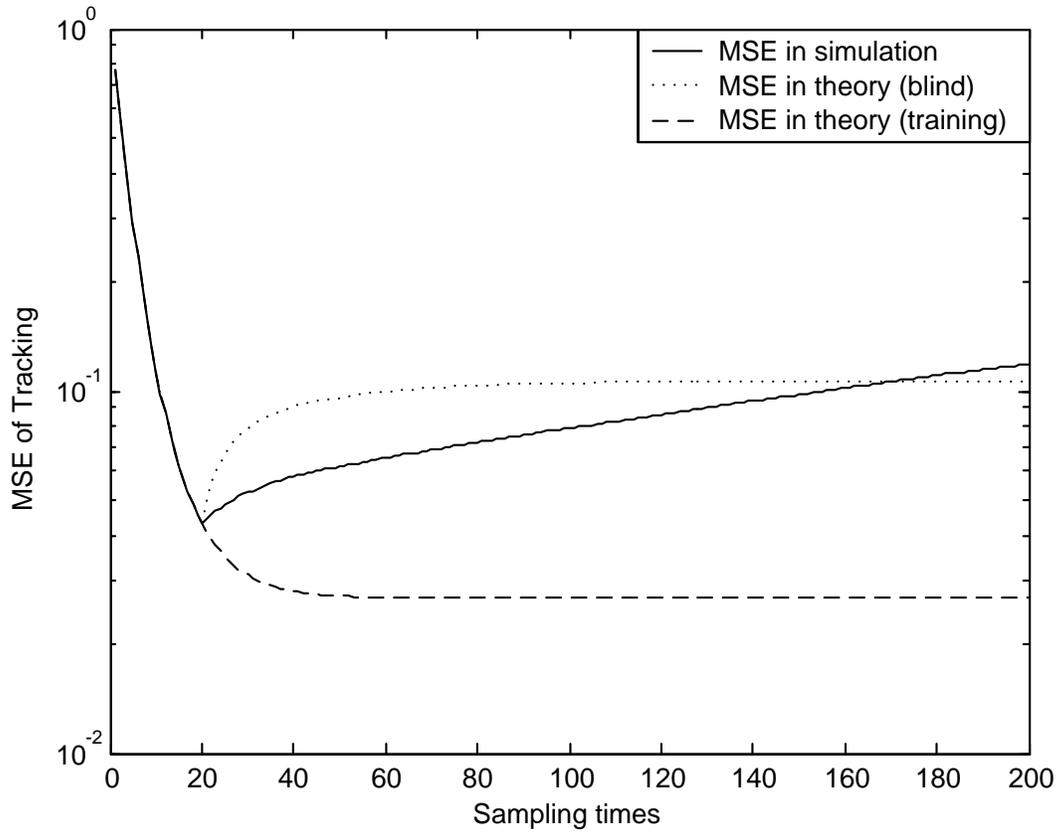

Fig. 9. Comparison of the MSE of Tracking in simulation and analysis where $f_D T = 0.01$ and $4 \times 4$ channel.

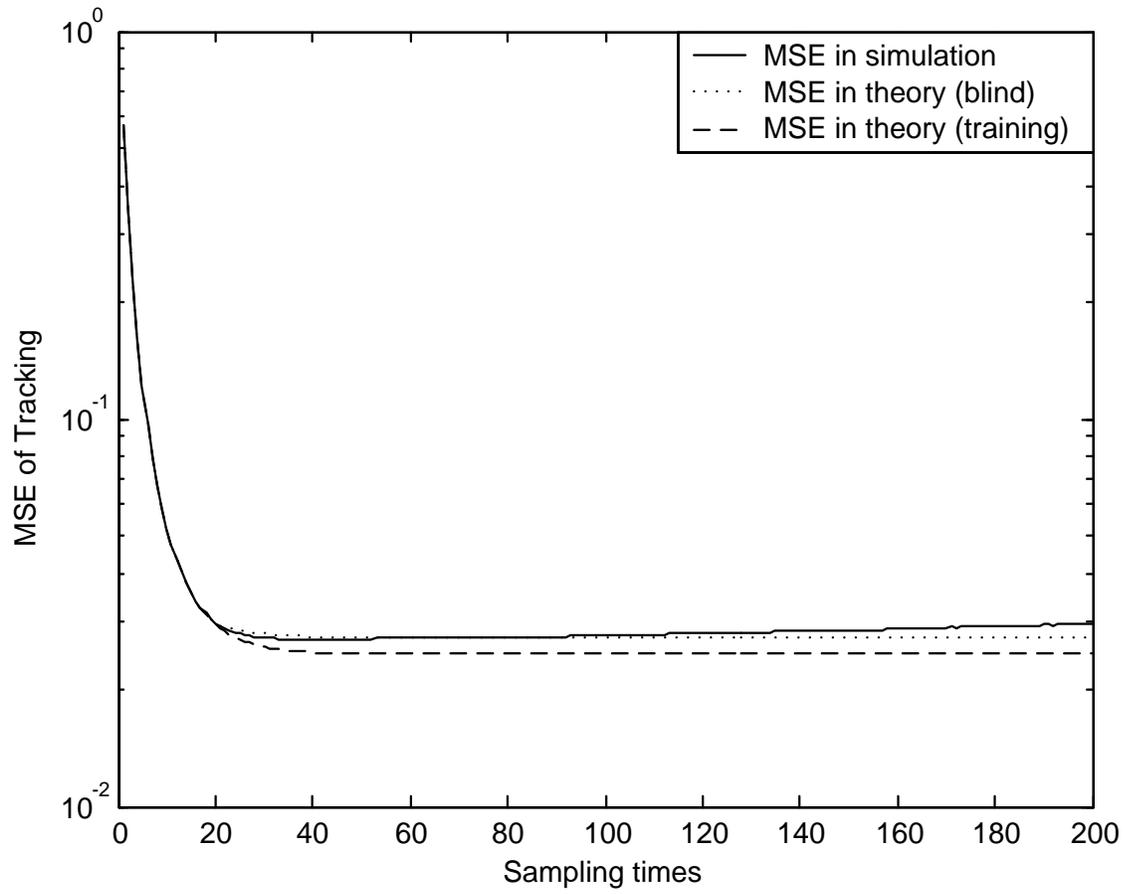

Fig. 10. Comparison of the MSE of Tracking in simulation and analysis where $f_D T$ =0.01 and 2×4 channel.